\documentclass{appolb}
\usepackage{graphicx}

\begin{document}
\title{Overview of heavy-flavour measurements in ALICE%
\thanks{Presented at Excited QCD 2020}%
}
\author{L.V.R. van Doremalen on behalf of the ALICE Collaboration
\address{Utrecht University, The Netherlands}
\\
{
}
}
\maketitle
\begin{abstract}
ALICE is devoted to the study of the properties of the Quark-Gluon Plasma (QGP). This state of matter is created in ultra-relativistic heavy-ion collisions at the LHC. Heavy quarks are considered effective probes of the QGP since, due to their large masses, they are produced in hard scattering processes and experience the full evolution of the hot and dense medium while interacting with its constituents. The heavy-quark measurements provide insights on processes like in-medium energy loss and hadronization. Measurements in proton-proton collisions provide the baseline for interpreting heavy-ion collision results and constitute an excellent test of pQCD calculations. In addition, proton-nucleus collisions allow separating cold nuclear matter effects from those due to the deconfined strongly interacting matter created in heavy-ion collisions. In this contribution, an overview of recent ALICE results for open heavy flavours, quarkonia, and heavy-flavour jets is presented. 
\end{abstract}
  
\section{Introduction}
At the extremely high temperatures reached in heavy-ion collisions, a phase-transition occurs from ordinary nuclear matter to a QGP state in which quarks and gluons are not confined into hadrons. The quark formation time during the collision is proportional to the inverse of the quark mass \cite{MassDep}. Therefore, heavy quarks are generated early during the collision and can experience the full evolution of the medium \cite{qgptime}. The quarks lose energy while moving through the medium by collisional and radiative processes. This energy loss is expected to depend on the path length, the QGP density,the parton colour charge (Casimir factor), and the quark mass (dead-cone effect) \cite{deadcone,charmbeauty}. Because of this, the following energy loss hierarchy is expected: $\Delta E_\mathrm{loss}$(g) $>$ $\Delta E_\mathrm{loss}$(u,d) $>$ $\Delta E_\mathrm{loss}$(c) $>$ $\Delta E_\mathrm{loss}$(b). 

The nuclear modification factor ($R_\mathrm{AA}$) quantifies the medium effects that affect the heavy quarks when they traverse the medium. This factor, defined as $$R_\mathrm{AA} = \frac{1}{\langle N^\mathrm{AA}_{coll}\rangle} \frac{\mathrm{d}N^\mathrm{AA} / \mathrm{d}p_\mathrm{T}}{\mathrm{d}N^\mathrm{pp} / \mathrm{d}p_\mathrm{T}},$$ is obtained from the ratio of the transverse-momentum-differential yields measured in Pb—Pb and pp collisions. The scaling factor $\langle N^\mathrm{AA}_{coll}\rangle$ represents the average number of binary nucleon-nucleon collisions in Pb--Pb collisions for a given centrality interval. If heavy quarks do not lose energy in the medium $R_\mathrm{AA} = 1$, while it drops below unity if they do. Heavy quarks are also expected to be affected by the collective motion of the medium. This gives rise to an anisotropic flow usually described by the components of a Fourier expansion of the azimuthal distribution of the outgoing particles. The second coefficient of this expansion is called elliptic flow ($v_2$). 

\begin{figure}[t!]
\begin{minipage}[b]{6cm}
\centerline{
\includegraphics[width=\textwidth]{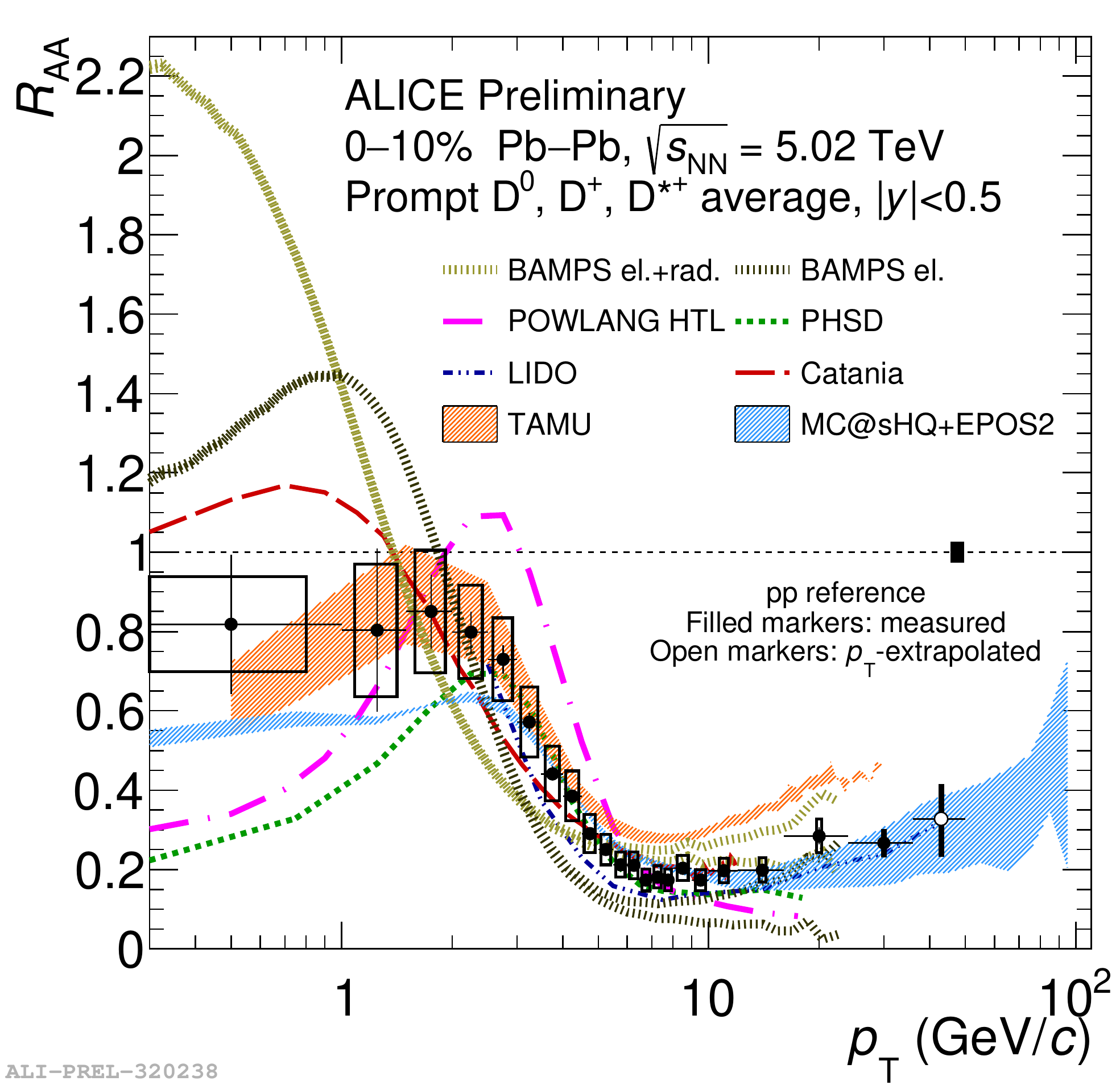}}
\end{minipage}
\begin{minipage}[b]{6cm}
\centerline{
\includegraphics[width=\textwidth]{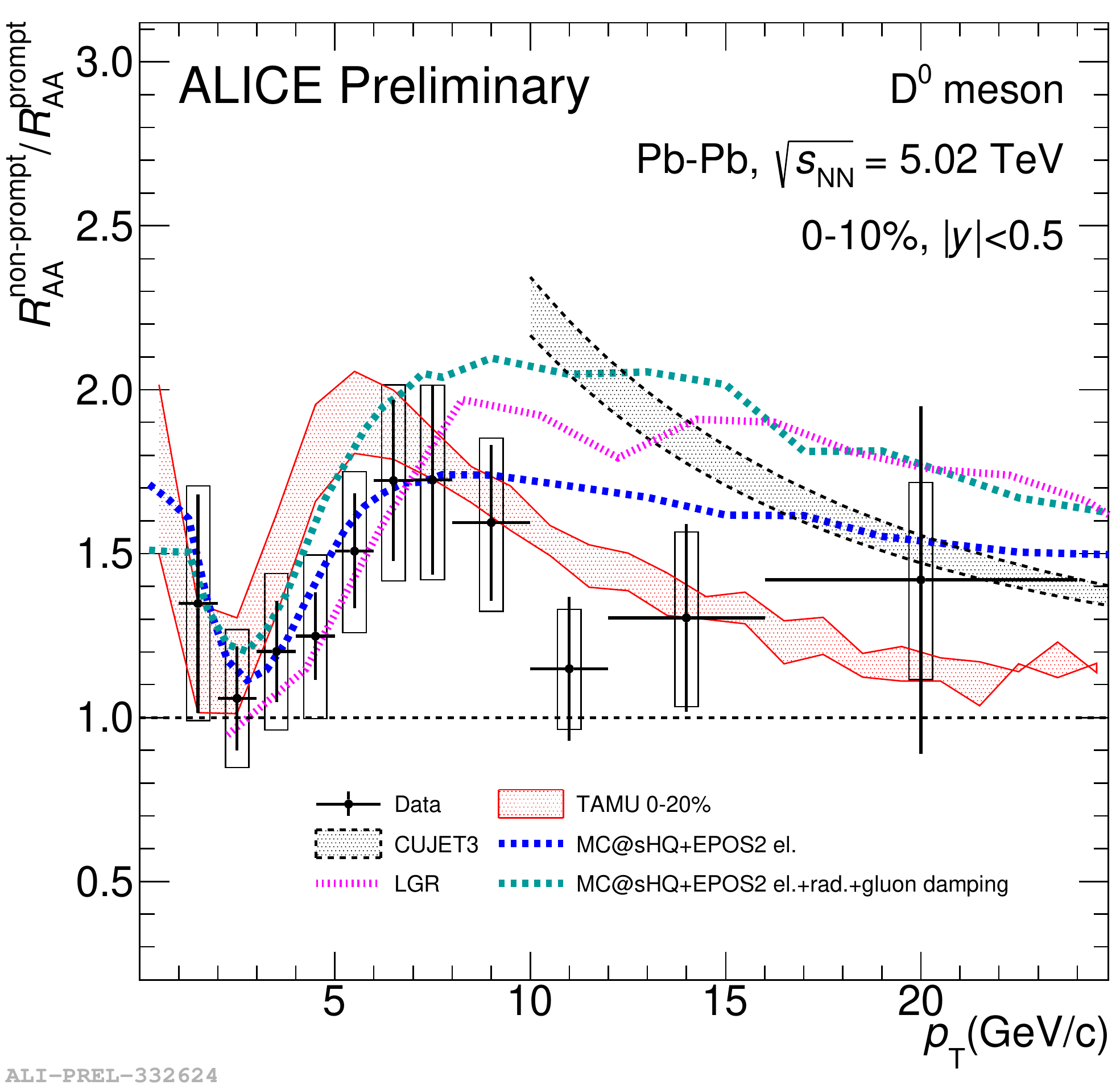}}
\end{minipage}
\caption{Left: $R_\mathrm{AA}$ of non-strange D mesons in central Pb--Pb collisions compared with theoretical calculations. Right: Ratio of $R_\mathrm{AA}$ of non-prompt D$^0$ mesons over the $R_\mathrm{AA}$ of prompt D$^0$ mesons. The data is compared with models with different energy loss for charm and beauty. Copyright CERN, reused with permission.}
\label{Fig:1}
\end{figure}

\section{Open heavy flavour}
The left panel in Fig. \ref{Fig:1} shows a comparison of the $R_\mathrm{AA}$ of non-strange D-mesons in central Pb--Pb collisions with theoretical calculations. The low momentum reach in central collisions allows setting stringent constraints on energy-loss models for central Pb--Pb collisions. Models without shadowing, like the BAMPS model \cite{bamps}, overestimate the $R_\mathrm{AA}$ spectrum at low $p_\mathrm{T}$. 

The models can be tested more rigorously by requiring a description of multiple observables, like $R_\mathrm{AA}$ and $v_2$, at the same time, over a wide momentum range, and in different centrality intervals \cite{dmesonRAA, dmesonFlow}. This shows that accurate modeling of data requires a combination of collisional and radiative energy loss, hadronization via coalescence, cold-nuclear-matter effects, and a realistic description of the medium evolution. 

The right panel shows the ratio of the $R_\mathrm{AA}$ of non-prompt D$^0$-mesons over the $R_\mathrm{AA}$ for prompt D$^0$-mesons. Prompt D$^0$-mesons, which come directly from the charm quarks produced in the initial collision, and non-prompt D$^0$-mesons, which are produced later by the decay of beauty hadrons, show a different $R_\mathrm{AA}$ at intermediate $p_\mathrm{T}$. Models with different energy loss for charm and beauty can describe within uncertainties  the ratio of non-prompt over prompt D$^0$-meson $R_\mathrm{AA}$. This is an indication that energy loss depends on the quark mass. 

\begin{figure}[t!]
\begin{minipage}[b]{6cm}
\centerline{
\includegraphics[width=\textwidth]{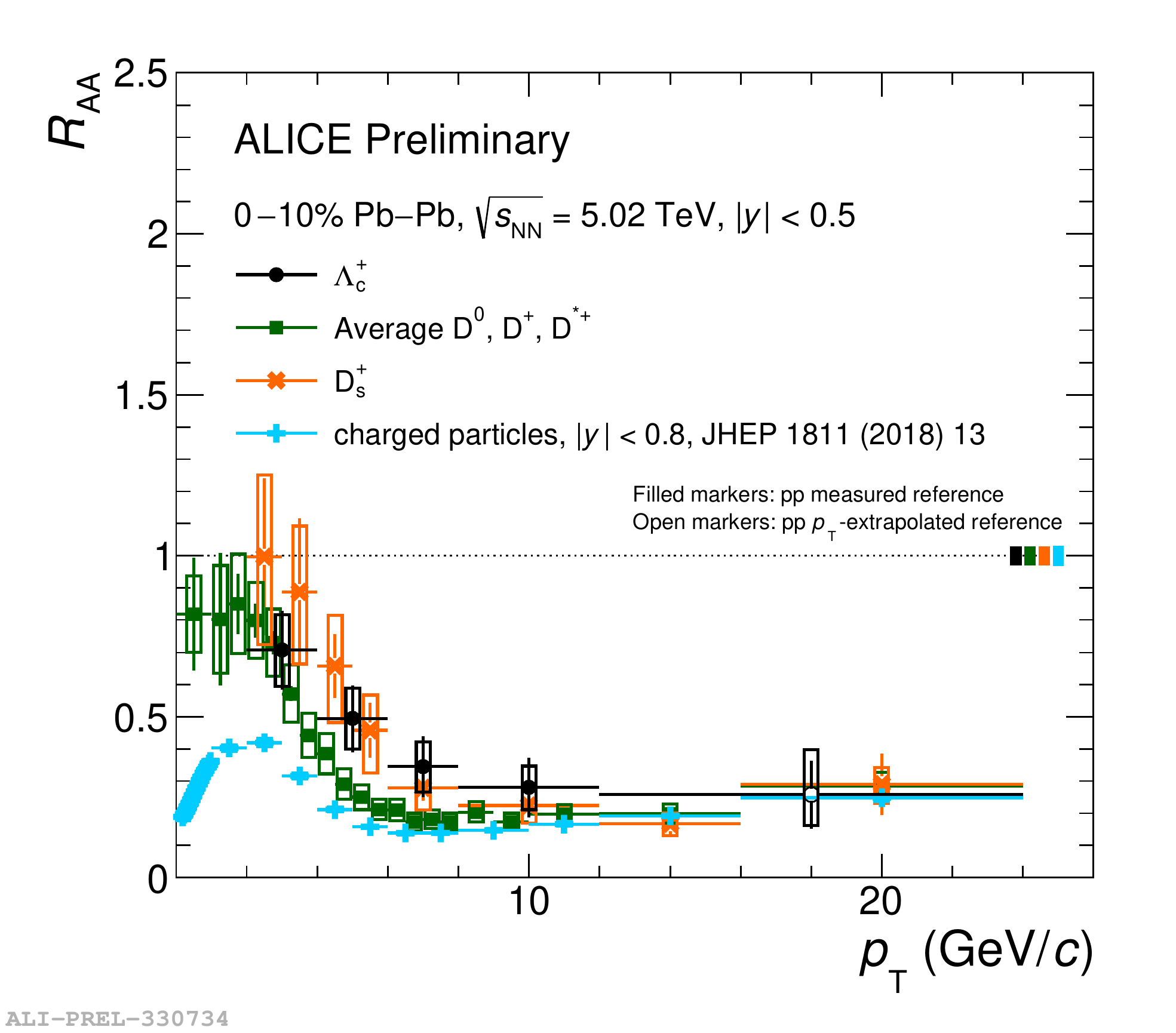}}
\end{minipage}
\begin{minipage}[b]{6cm}
\centerline{
\includegraphics[width=\textwidth]{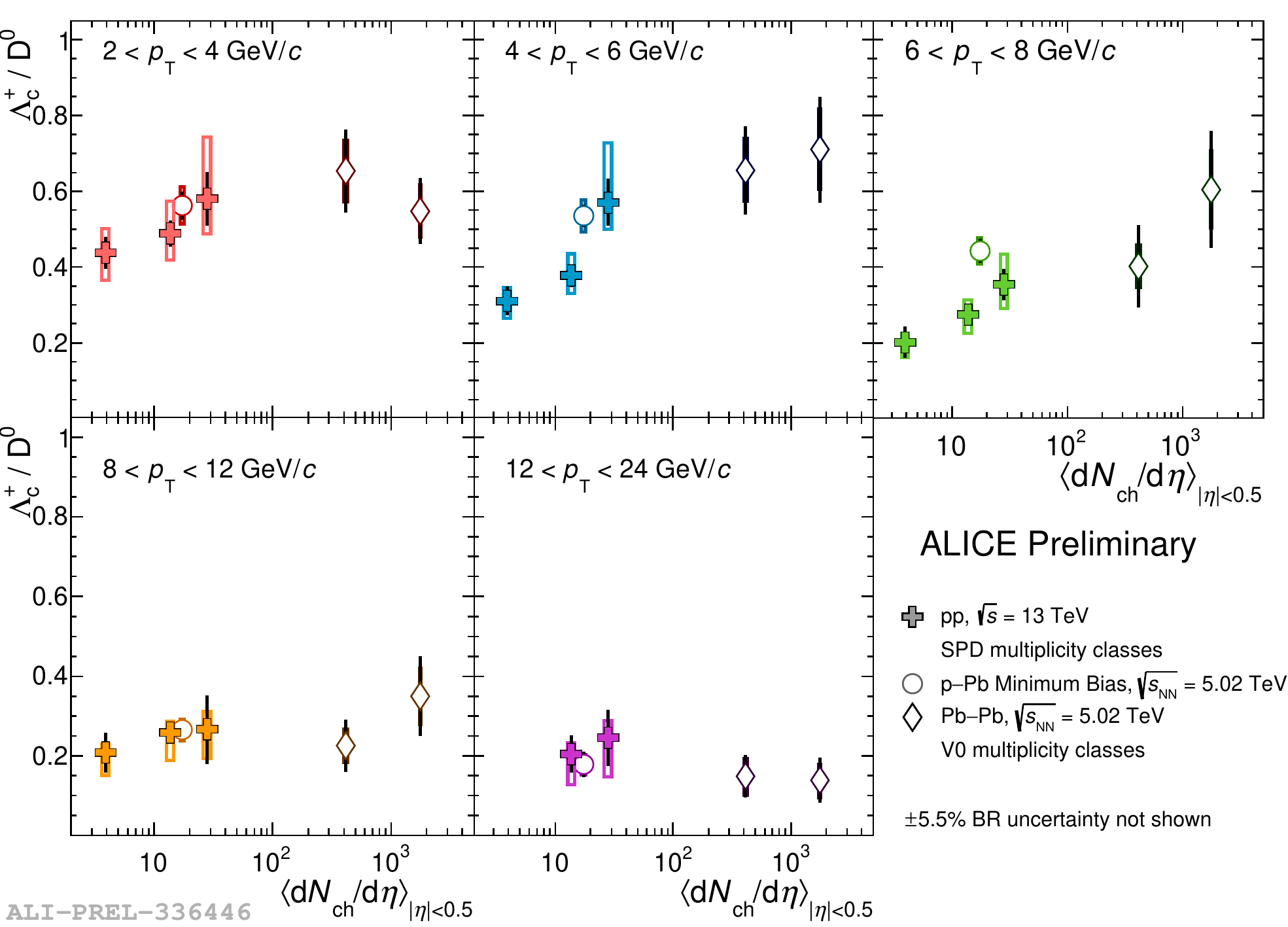}}
\end{minipage}
\caption{Left: $R_\mathrm{AA}$ in central Pb--Pb collisions for multiple types of particle species. Right: $\Lambda_c^+$  / D$^0$ ratio as a function of multiplicity for several $p_\mathrm{T}$ intervals.  Copyright CERN, reused with permission.}
\label{Fig:2}
\end{figure}

The left panel in Fig. \ref{Fig:2} shows the $R_\mathrm{AA}$ for different particle species with a hierarchy that is consistent with the expected difference in energy loss for charm versus light-flavour and gluons. Strange D-mesons and $\Lambda_{c}$ baryons show a hint of lower suppression, compared to non-strange D-mesons, that may point at recombination effects. "Models that include hadronization via coalescence reproduce D$_\mathrm{S}$ data within uncertainties. 

The right panel in Fig. \ref{Fig:2} shows the $\Lambda_c^+$ / D$^0$ ratio as a function of multiplicity in pp, p--Pb, and Pb--Pb collisions for several $p_\mathrm{T}$ intervals. This ratio shows an enhancement at low $p_\mathrm{T}$ compared to e$^{+}$e$^{-}$ collider measurements in which $\Lambda_c^+$ / D$^0$ $\approx 0.1$ \cite{epref}. The multiplicity dependence of the $\Lambda_c^+$  / D$^0$ ratio shows that the enhancement remains higher than electron-positron collider measurements even for low-multiplicity pp collisions, suggesting that charm-quark recombination with quarks from the surrounding hadronic environment may already occur in small systems.

\begin{figure}[t!]
\begin{minipage}[b]{6cm}
\centerline{
\includegraphics[width=\textwidth]{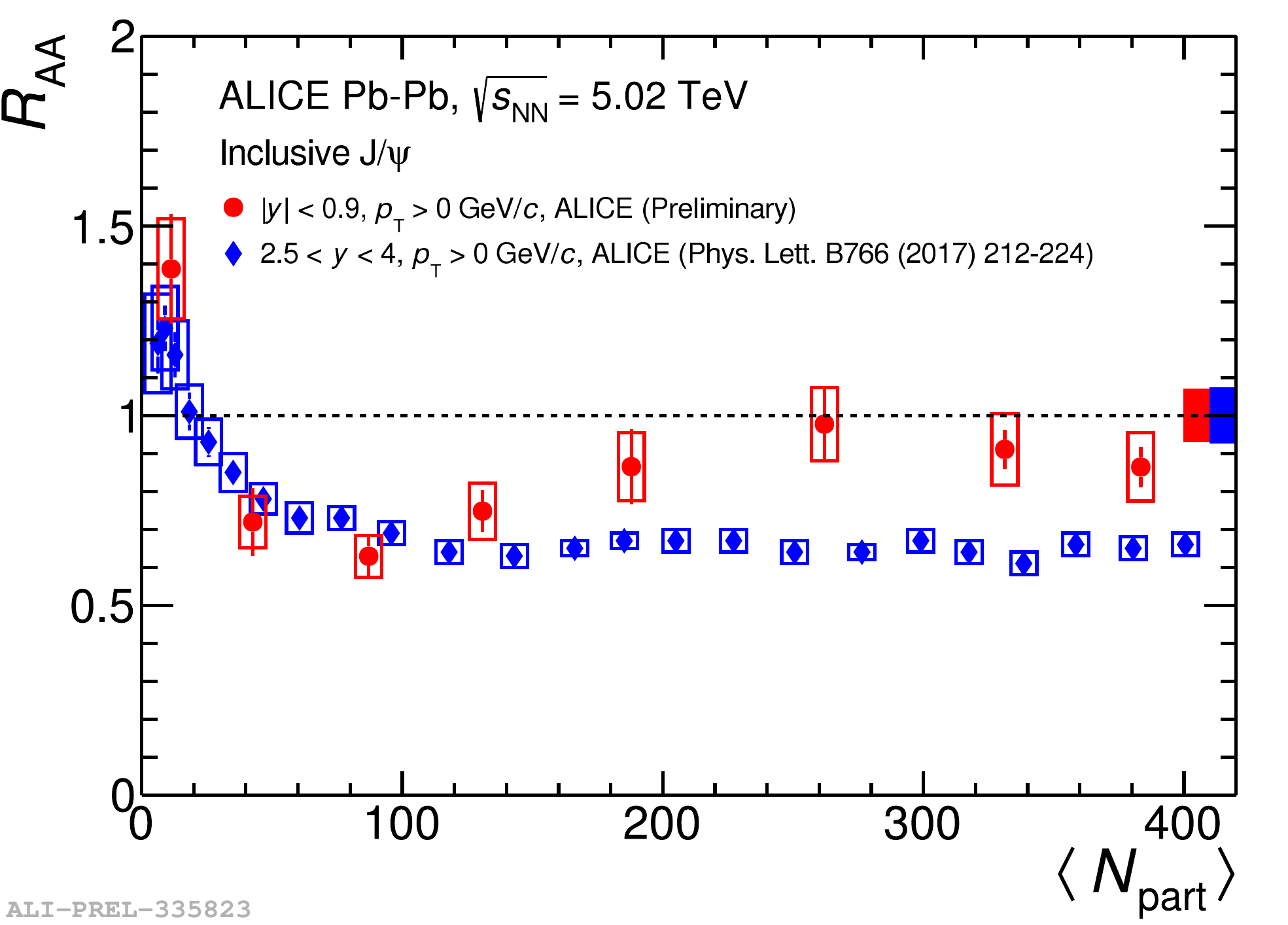}}
\end{minipage}
\begin{minipage}[b]{6cm}
\centerline{
\includegraphics[width=\textwidth]{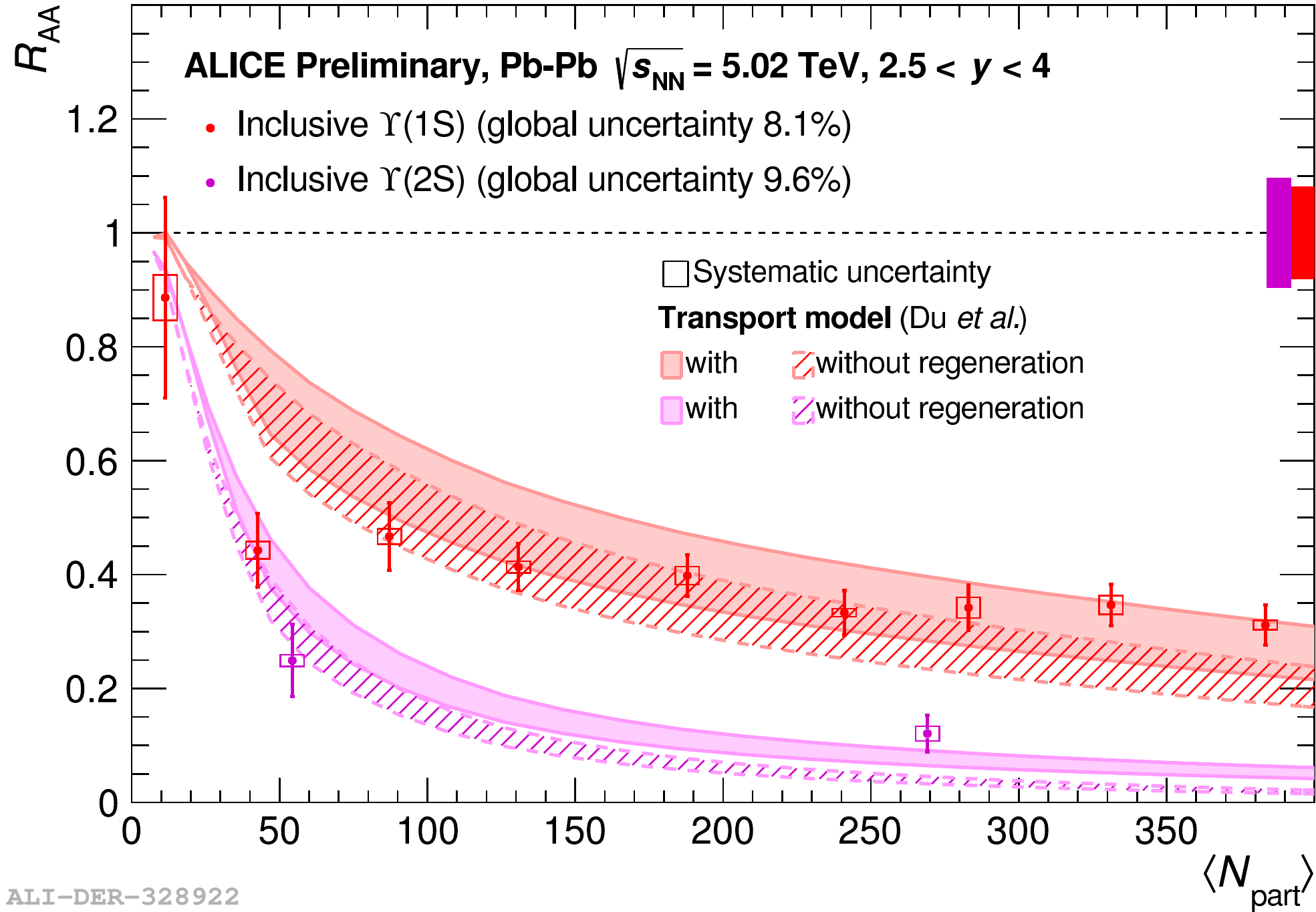}}
\end{minipage}
\caption{Left: $R_\mathrm{AA}$ as a function of multiplicity for inclusive J/$\psi$ in two rapidity intervals. Right: $R_\mathrm{AA}$ as a function of $\langle N_\mathrm{part} \rangle$ for two $\Upsilon$ states along with model predictions. Copyright CERN, reused with permission.}
\label{Fig:3}
\end{figure}

\section{Quarkonium}
At high temperatures colour screening in the QGP results in the suppression of quarkonium production \cite{quarkonium}. Different quarkonium states have different binding energies, which results in the expectation of a sequential melting of states when colliding nuclei at higher energies \cite{melting}. On the other hand, the c$\bar{\mathrm{c}}$ multiplicity increases at higher collision energies. This leads to the expectation of an enhancement of quarkonia production via recombination at hadronization.

The left panel of Fig. \ref{Fig:3} shows the $R_\mathrm{AA}$ as a function of multiplicity for inclusive J/$\psi$-mesons in two rapidity intervals. This $R_\mathrm{AA}$ measurement has a significantly improved precision and $p_\mathrm{T}$ reach compared to previous measurements \cite{improved}. At higher multiplicities the $R_\mathrm{AA}$ at midrapidity is higher than at forward rapidity. This observation may suggest that recombination effects are stronger at midrapidity, where the charm-quark density is higher.

The centrality dependence of the $R_\mathrm{AA}$ is shown in the right panel of Fig. \ref{Fig:3}. The data show a slight bottomonium centrality dependence and match well with the model predictions \cite{du}. A stronger suppression of $\Upsilon$(2S) than $\Upsilon$(1S) is observed.

\begin{figure}[t!]
\begin{minipage}[b]{6cm}
\centerline{
\includegraphics[width=\textwidth]{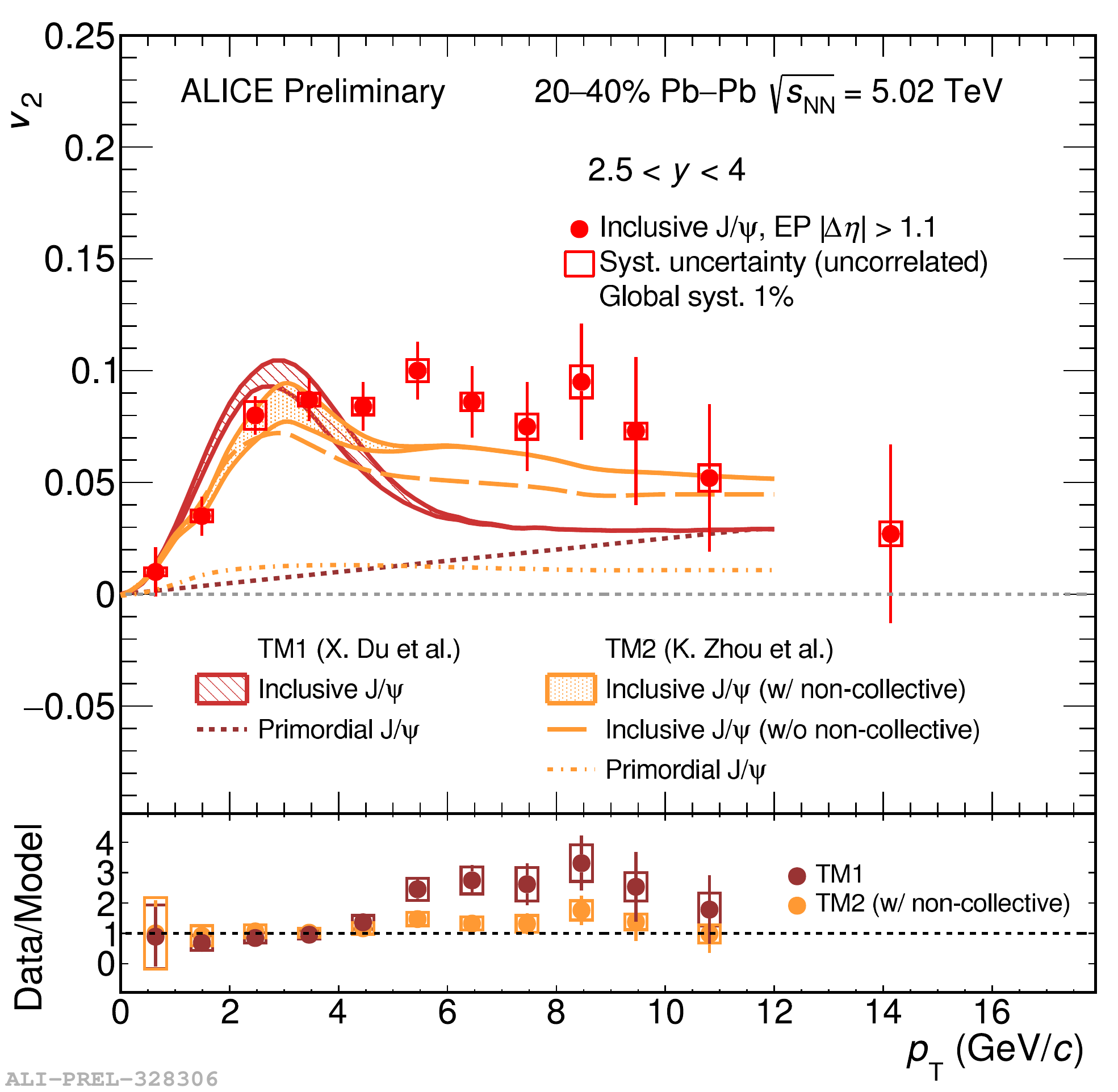}}
\end{minipage}
\begin{minipage}[b]{6cm}
\centerline{
\includegraphics[width=\textwidth]{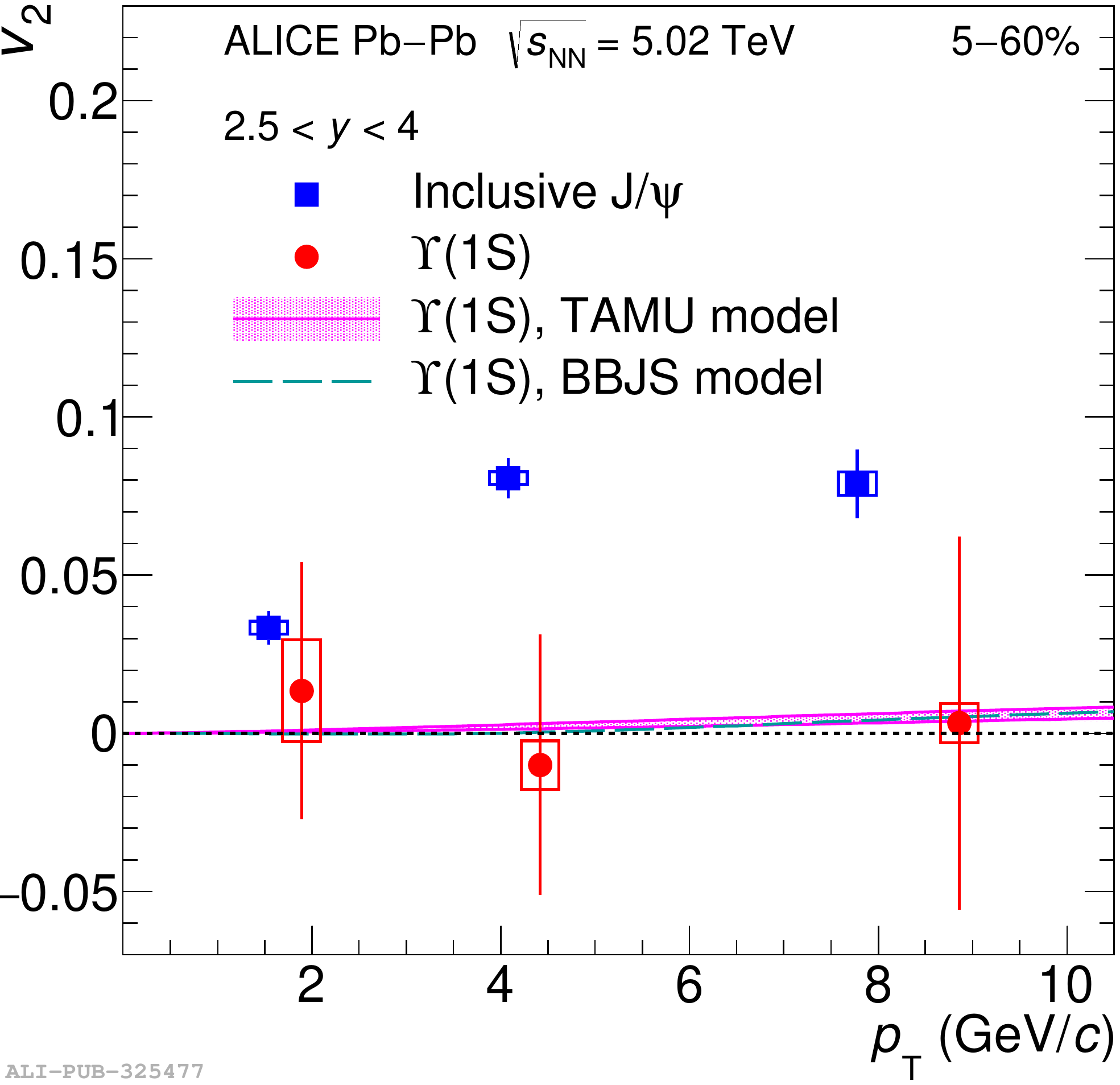}}
\end{minipage}
\caption{Left: $v_2$ as a function of $p_\mathrm{T}$ for inclusive J/$\psi$. Right: $\Upsilon$(1S) $v_2$ as a function of $p_\mathrm{T}$ compared with inclusive J/$\psi$ $v_2$ and different models \cite{Yflow}. Copyright CERN, reused with permission.}
\label{Fig:4}
\end{figure}

For J/$\psi$-mesons, measurements show a positive $v_2$ in a large $p_\mathrm{T}$ range at forward rapidity. This is illustrated in the left panel of Fig. \ref{Fig:4}. The bottomonium $v_2$ is consistent with zero, however more data are needed for a conclusive interpretation on the difference between J/$\psi$ and bottomonium $v_2$

\section{Heavy-flavour jets}
Jets originate from hard parton-parton interactions. In ALICE heavy-flavour tagged jets are measured down to low jet $p_\mathrm{T}$ (5 GeV/$c$). The study of jets provides experimental data for gluon-to-hadron fragmentation functions and gluon PDF at low $x$. The study of jet quenching provides additional information to further characterise parton energy loss in the QGP.

\begin{figure}[t!]
\begin{minipage}[b]{6cm}
\centerline{
\includegraphics[width=\textwidth]{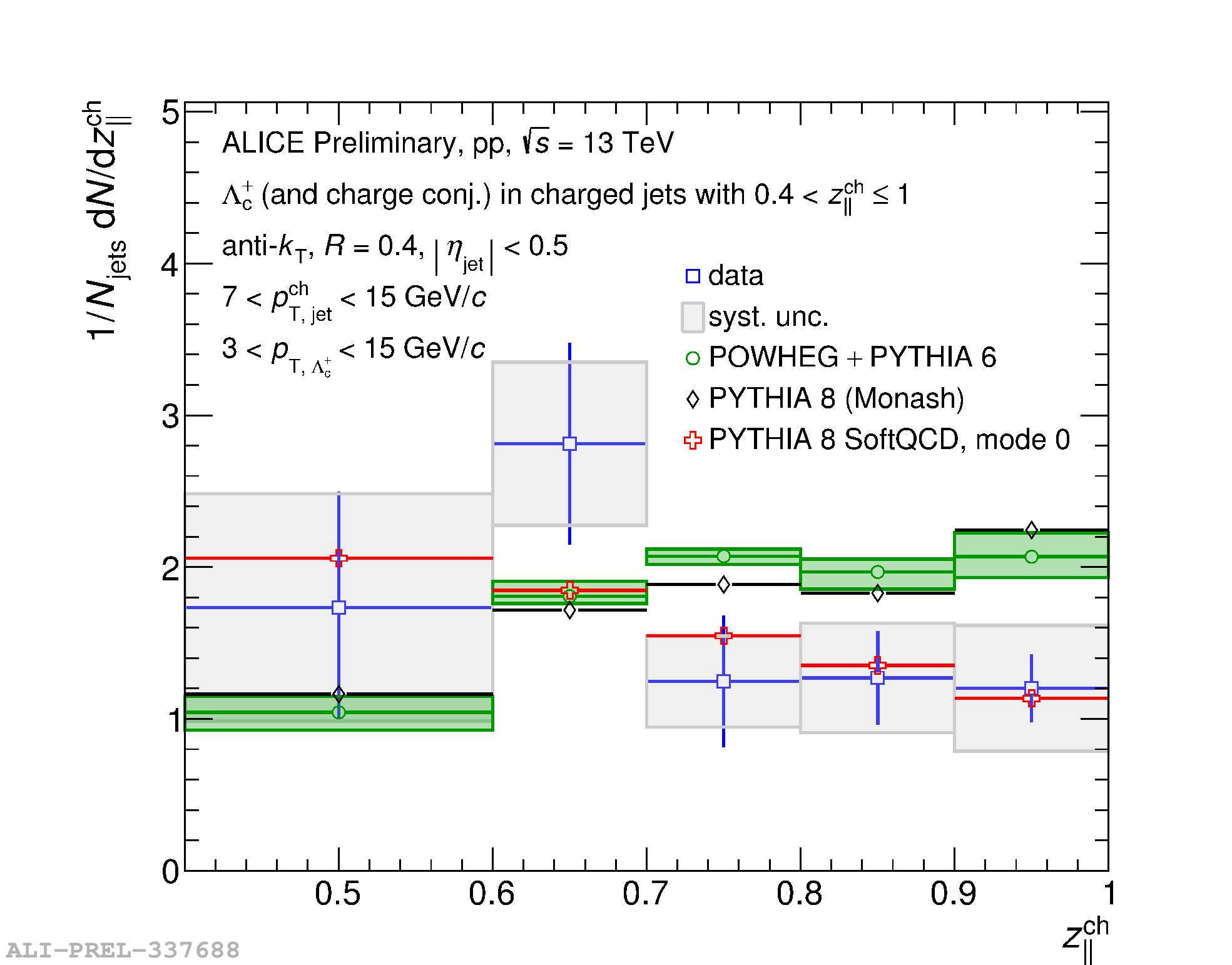}}
\end{minipage}
\hspace{0.1cm}
\begin{minipage}[b]{6cm}
\centerline{
\includegraphics[width=\textwidth]{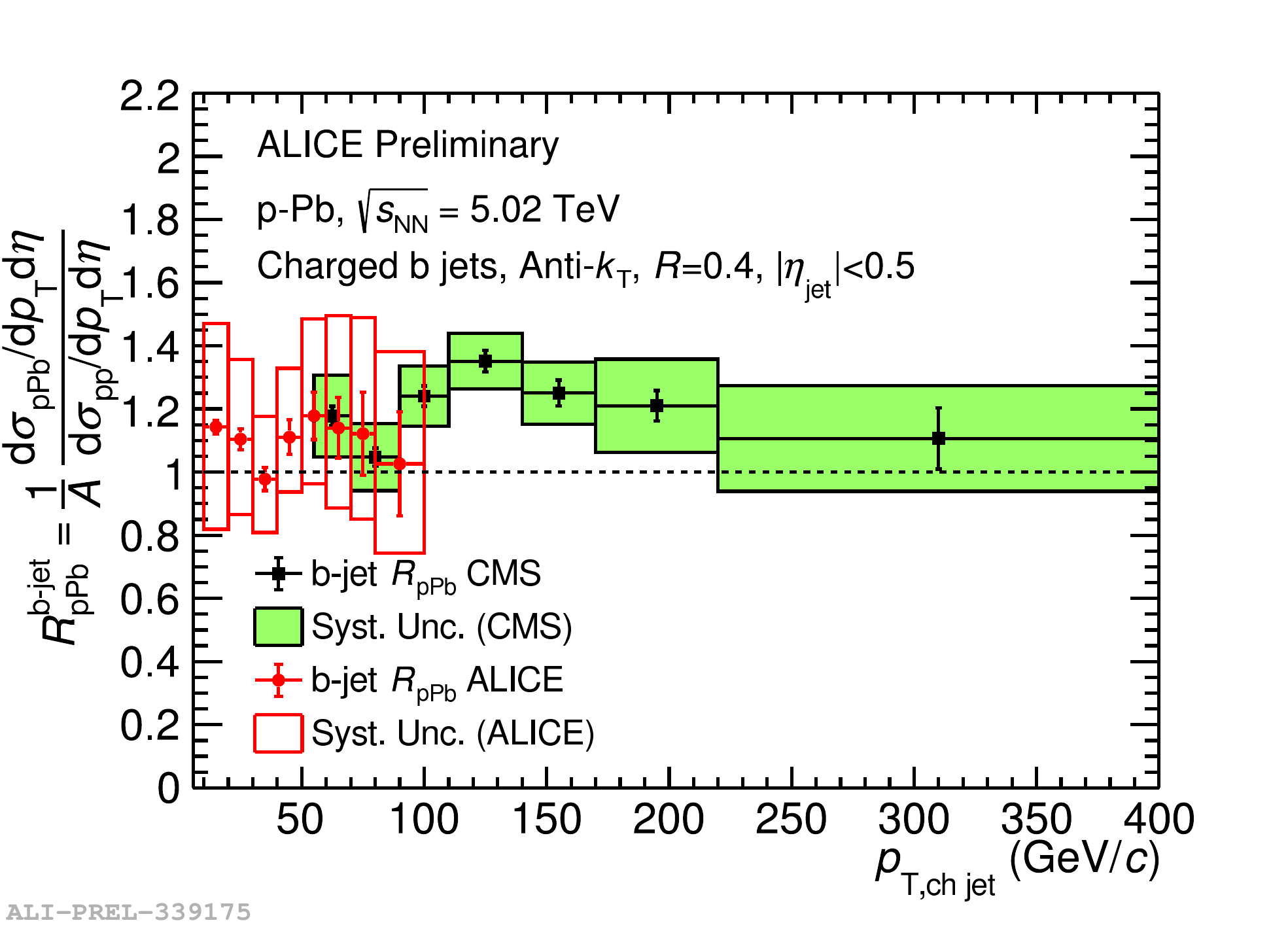}}
\end{minipage}
\caption{Left: probability density distribution of the parallel jet momentum fraction (z$_{||}^{ch}$) for $\Lambda_c^+$-tagged jets compared to expectations from Monte Carlo generators. Right: $R_\mathrm{pPb}$ for B-tagged jets with a comparison of measurements by ALICE and CMS \cite{CMSbjet}. Copyright CERN, reused with permission.}
\label{Fig:5}
\end{figure}

Fig. \ref{Fig:5} shows the first measurement of the $\Lambda_c^+$ probability density distribution of the parallel jet momentum fraction (z$_{||}^{ch}$) compared to data. The Pythia 8 SoftQCD model has the best agreement with data.

 Jets with beauty hadrons were reconstructed exploiting the displaced impact parameter of b-hadron decay tracks to the primary vertex. The observed yields are consistent with POWHEG. The nuclear modification factor in p--Pb ($R_\mathrm{pPb}$) for B-tagged jets is shown in the right panel of Fig. \ref{Fig:5}. No cold-nuclear-matter effects are observed within uncertainties using B-tagged jets.

\end{document}